\documentstyle[11pt,newpasp,psfig,epsfig,twoside]{article}
\def\kmsmpc{\,{\rm km\,s^{-1}\,Mpc^{-1}}}

\def\eblunits{\,{\rm nW\,m^{-2}\,sr^{-1}}}

\def\msun{\,{\rm M_\odot}}
\def\mden{\,{\rm M_\odot\,Mpc^{-3}}}
\def\lden{\,{\rm L_\odot\,Mpc^{-3}}}
\def\sfrd{\,{\rm M_\odot\,yr^{-1}\,Mpc^{-3}}}

\def\etal{{et al.\ }}

\def\spose#1{\hbox to 0pt{#1\hss}}
\def\lta{\mathrel{\spose{\lower 3pt\hbox{$\mathchar"218$}}
     \raise 2.0pt\hbox{$\mathchar"13C$}}}
\def\gta{\mathrel{\spose{\lower 3pt\hbox{$\mathchar"218$}}
     \raise 2.0pt\hbox{$\mathchar"13E$}}}

\begin{document}
\title{Extragalactic Background Light, MACHOs, and the Cosmic Stellar Baryon Budget}
\author{Piero Madau}
\affil{Department of Astronomy and Astrophysics, University of California,
Santa Cruz, CA 95064, USA}
\author{Francesco Haardt}
\affil{Dipartimento di Scienze, Universit\'a dell'Insubria, via Lucini 3, 
Como, Italy}
\author{Lucia Pozzetti}
\affil{Osservatorio Astronomico di Bologna, Via Ranzani 1, 40127 Bologna, 
Italy}

\begin{abstract}
\noindent The optical/far--IR extragalactic background light (EBL) from both 
resolved and unresolved 
extragalactic sources is an indicator of the total 
luminosity of cosmic structures, as the cumulative emission from young and evolved 
galactic systems, as well as from active galactic nuclei (AGNs), is recorded 
in this radiation. This is a brief review of some of the implications of the
observed brightness of the night sky for the stellar mass density and
average metallicity of the universe today, and of the possible contribution 
of MACHO progenitors and QSOs to the EBL.
Assuming a Salpeter initial mass function with a cutoff below $0.6\,\msun$,
a lower limit of $\Omega_{g+s}h^2>0.0015\,I_{60}$ can be derived to the visible 
(recycled gas $+$ stars) mass density required to generate an 
EBL at a level of $I_{\rm EBL}=60\,I_{60}\,\eblunits$: our latest, `best--guess' 
estimate is $\Omega_{g+s} h^2 \approx 0.0023\,I_{60}$, which implies a mean 
metallicity at the present--epoch of $y_Z\Omega_{g+s}/\Omega_b\approx 0.2\,Z_\odot$. 
If massive dark halos around spiral galaxies are partially composed of faint, old
white dwarfs, i.e. if a non--negligible fraction ($\sim$ a few percent) of the 
nucleosynthetic baryons is locked in the remnants (MACHOs) of intermediate--mass 
stars forming at very high redshifts, then the bright early phases of such halos 
should contribute significantly to the observed EBL.
Assuming a standard black hole accretion model for quasar activity and using 
recent observations of the quasar population and new synthesis models 
for the cosmic X--ray background, we estimate a present mass density of QSO 
remnants of $\rho_{\rm BH}\approx 3\times 10^5\,\mden$ for a 10\% efficiency of 
accreted mass--to--radiation conversion.
The quasar contribution to the brightness of the night sky is $I_{\rm QSO}\approx 
2\,\eblunits$.
\end{abstract} 

\section{Introduction}

Observational studies of the distant universe are
undergoing a revolution brought about by breakthroughs achieved with the
{\it HST}, {\it Keck}, and {\it JCMT} telescopes, and
the {\it ISO} satellite.
The remarkable progress in our understanding of faint
galaxy data made possible by the combination of {\it HST} deep imaging
and ground--based spectroscopy has
permitted in the last few years to shed new light on the evolution of
the stellar birthrate in the universe, to identify the epoch where most of
the optical extragalactic background light was produced, and to set important
contraints on galaxy evolution scenarios.
The use of novel instruments and observational techniques has led to
the measurement of the abundance and clustering of actively star--forming 
objects at redshift $3-4$ and to the discovery of galaxies and QSOs at redshift
in excess of 5, when the universe was less than 6\% of its current age.

The explosion in the quantity of information available on the high--redshift
universe at optical wavelengths has been complemented by the detection
by DIRBE and FIRAS onboard the {\it COBE} satellite of the far--IR/sub--mm
background -- which has revealed that a significant fraction of the energy 
released by stellar nucleosynthesis is re--emitted as thermal radiation by 
dust -- and by theoretical progress made in understanding how baryons 
follows
the dynamics dictated by dark matter halos until radiative, hydrodynamic, and 
star formation processes take over.
While ongoing studies with {\it Chandra} and {\it XMM--Newton} may be
discovering a new population of highly absorbed, dusty AGNs (Type 2 QSOs),    
of importance for galaxy formation models also appear the
recent findings of a strong link between the masses of black holes in 
the nuclei of nearby galaxies and the properties of the host stellar bulges, 
and the suggestion from Magellanic Cloud microlensing experiments and 
systematic 
proper--motion surveys that a non--negligible fraction of the dark matter in the 
Galactic halo may be tied up in very old, cool white dwarfs. The underlying 
goal of all these efforts is to understand the growth of structures,
the internal properties of galaxies and their evolution, and ultimately
to map the star formation and supermassive black hole accretion histories 
of the universe 
from the end of the `dark age' to the present epoch. The implications of
the observed energy density of the EBL for the cosmic stellar and metal 
budget, and the quasar contribution to the brightness of the night sky, will 
be the subject of this talk.  
Unless otherwise stated, an Einstein-de Sitter (EdS) cosmology with 
$\Omega_M=1$, $\Omega_\Lambda=0$, and $H_0=100\,h\,\kmsmpc$ will be adopted 
in the following. 

\section{The brightness of the night sky}

The logarithmic slope of the galaxy number--magnitude relation is a remarkably 
simple cosmological probe of the history of stellar birth in galaxies, as 
it must drop below 0.4 to yield a finite value for the EBL.
This appears to be the case in all seven $UBVIJHK$ optical bandpasses, i.e. the 
light from resolved galaxies has converged from the UV to the near--IR (Madau
\& Pozzetti 2000, hereafter MP). The flattening at faint 
apparent magnitudes cannot be due to the reddening of distant sources as 
their Lyman break gets redshifted into the blue passband, since the fraction 
of Lyman--break galaxies at (say) $B\approx 25$ is small.
Moreover, an absorption--induced loss of sources cannot explain the similar
change of slope of the galaxy counts observed in the $V,I,J,H,$ and $K$ bands.
While this suggests that the surface density of 
optically luminous galaxies is leveling off beyond $z\sim 1.5$, 
one should note that different algorithms used for `growing' the
photometry beyond the outer isophotes of galaxies may significantly change
the magnitude of faint objects. According to Bernstein \etal (2000), roughly 
50\% of the flux from resolved galaxies with $V>23$ mag lie outside the 
standard--sized apertures used by photometric packages. An extragalactic
sky pedestal created by the overlapping wings of resolved galaxies may 
contribute to the sky level, and would be undetectable except 
by absolute surface photometry. Also, at faint 
magnitude levels, distant objects which are brighter than the nominal depth of 
the catalog may be missed due to the $(1+z)^4$ dimming factor.
All these systematic errors are inherent in faint--galaxy photometry;
as a result,  estimates of the integrated fluxes from resolved galaxies 
should be strictly considered as lower limits (see Pozzetti \& Madau, this
volume). 

The spectrum of the optical EBL 
is shown in Figure 1, together with the recent results from {\it COBE}.
The value derived by integrating the galaxy counts down to very
faint magnitude levels (because of the flattening of the number--magnitude 
relation most of the contribution to the 
optical EBL comes from relatively bright galaxies) 
implies a lower limit to the EBL intensity in the 0.2--2.2 $\mu$m 
interval of $I_{\rm opt}\approx 15\,\eblunits$. Including the tentative 
detections  at 2.2 and 3.5 $\mu$m by Gorjan \etal (2000) (see also 
Dwek \& Arendt 1998) would boost $I_{\rm opt}$ to $\gta 20\,\eblunits$. 
Recent direct measurements of the EBL at 3000, 5500, and 8000 \AA\ 
from absolute surface photometry lie between a 
factor of 2.5 to 3 higher than the integrated light from galaxy counts, with 
an uncertainty that is largely due to systematic rather than 
statistical error (Bernstein \etal 2000). Applying this correction factor to 
the range 
3000--8000 \AA\ gives a total optical EBL intensity in the range $25-30\,
\eblunits$. This could become $\sim 45\,\eblunits$ if the same correction 
holds also in the near--IR (Gorjan \etal 2000). The {\it COBE}/FIRAS (Fixsen 
\etal 1998) 
measurements yield $I_{\rm FIR}\approx 14\,\eblunits$ in the 125--2000 $\mu$m 
range. When combined with the DIRBE (Hauser \etal 1998) 
points at 140 and 240 $\mu$m, one gets a far--IR 
background intensity of $I_{\rm FIR}(140-2000\,\mu{\rm m})\approx 
20\,\eblunits$.
The detection with DIRBE of a FIR signal in excess of the expected zodiacal and
Galactic emission by Finkbeiner \etal (2000), if confirmed, would imply
an integrated EBL in the window 45--2000 $\mu$m of $\gta 40\,\eblunits$. 
The residual emission in the 3.5 to 140 $\mu$m region is poorly known, but it 
is likely to exceed $10\,\eblunits$ (Dwek \etal 1998). 

A `best--guess' estimate 
of the total EBL intensity observed today appears to be
\begin{equation}
I_{\rm EBL}=60\pm 20\,\eblunits. 
\end{equation}
In the following, we will adopt a reference value for the background light 
associated with star formation activity
over the entire history of the universe of $I_{\rm EBL}=60\,I_{60}\eblunits$.  

\section{The stellar mass density today}

With the help of some simple stellar population synthesis tools we can now 
set a lower limit to the total stellar mass density 
that produced the observed EBL, and constrain 
the cosmic history of star birth in galaxies. One of the most serious 
uncertainties in this calculation is the lower cutoff, usually treated as a 
free parameter, of the initial mass function (IMF).
Observations of M subdwarfs stars with the {\it HST} have recently
shed some light on this issue, showing that the IMF in the Galactic disk can 
be represented analytically
over the mass range $0.1<m<1.6$ (here $m$ is in solar units) by $\log 
\phi(m)={\rm const} -2.33 \log m -1.82(\log m)^2$ (Gould \etal 1996, hereafter GBF).
For $m>1$ this 
mass distribution agrees well with a Salpeter function. A shallow mass 
function below $1\,M_\odot$ has also been 
recently measured in the Galactic bulge (Zoccali \etal 2000) and in globular
clusters (Paresce \& De Marchi 2000). Observations of normal 
Galactic star--forming regions also show 
some convergence in the basic form of the IMF at intermediate and high masses, 
a power--law slope that is consistent with the Salpeter value (Elmegreen 
1998; Massey 1998).
In the following we will use a `universal' IMF (shown in Figure 2) with the 
GBF form for $m<1$, matched to a Salpeter slope for $m\ge 1$; the mass 
integral of this function is 1.7 times smaller than that obtained by extrapolating 
a Salpeter function down to 
$0.1\, M_\odot$.\footnote{Since the bolometric light contributed by stars less 
massive than $1\,M_\odot$ is very small for a `typical' IMF, the use of a  
GBF mass function at low masses instead of Salpeter leaves the total 
radiated luminosity of a stellar population virtually unaffected.}

As shown in Figure 3, the {\it bolometric} 
luminosity as a function of age $\tau$ of a simple stellar population (a single 
generation of coeval, 
chemically homogeneous stars having total mass $M$, solar metallicity, and the
above IMF) can be well approximated by 
\begin{equation}
L(\tau)= \left\{\begin{array}{ll} 1200\,L_\odot {M\over M_\odot} & 
\mbox{$\tau\le 2.6\,$ Myr;} \\
0.7\,L_\odot {M\over M_\odot} \left({\tau\over 1\,{\rm Gyr}}\right)^{-1.25} 
& \mbox{$2.6\le \tau\le 100\,$ Myr;} \\
2.0\,L_\odot {M\over M_\odot} \left({\tau\over 1\,{\rm Gyr}}\right)^{-0.8} & 
\mbox{$\tau>
100\,$ Myr.} \end{array}
\right.
\end{equation}
Over a timescale of 13 Gyr -- the age of the universe for an EdS cosmology 
with $h=0.5$ -- about 1.3 MeV per stellar baryon will be 
radiated away. This number depends only weakly on the assumed metallicity of 
stars.  In a stellar system with arbitrary star formation rate per
comoving cosmological volume, $\dot \rho_s$, 
the bolometric emissivity at time $t$ is given by the convolution integral
\begin{equation}
\rho_{\rm bol}(t)=\int_0^t L(\tau)\dot \rho_s(t-\tau)d\tau.
\end{equation}
The total background light observed at Earth ($t=t_H$), 
generated by a stellar population with a formation epoch $t_F$, is 
\begin{equation}
I_{\rm EBL}={c\over 4\pi} \int_{t_F}^{t_H} {\rho_{\rm bol}(t)\over 1+z}dt,
\end{equation}
where the factor $(1+z)$ at the denominator is lost to cosmic expansion 
when converting from observed to radiated (comoving) luminosity density. 

To set a lower limit to the present--day mass density, $\Omega_{g+s}$, of 
recycled gas $+$ stars 
(in units of the critical density $\rho_{\rm crit}=2.8 
\times 10^{11}\,h^2\mden$), consider now a scenario where all stars 
are formed {\it instantaneously} at redshift $z_F$.
The background light that would be observed at Earth from such an event 
is shown in Figure 3 as a function of $z_F$ for $\Omega_{g+s}h^2=0.0008, 
0.0013, 0.0018$, corresponding respectively to 4, 7, and 9 percent of the 
nucleosynthetic baryon density, $\Omega_bh^2=0.0193\pm 0.0014$ (Burles \& 
Tytler 1998). Two main results are 
worth stressing here: (1) the time evolution of the luminosity radiated by a 
simple stellar population (eq. 2) makes the dependence of the 
observed EBL from $z_F$ much shallower than the $(1+z_F)^{-1}$ 
lost to cosmic expansion (see eq. 4), as the energy output from
stars is spread over their respective lifetimes; and (2) in order to generate 
an EBL at a level of $60\,I_{60}\,\eblunits$, one requires 
$\Omega_{g+s}h^2>0.0015\,I_{60}$ for an EdS universe with $h=0.5$, hence a mean 
mass--to--blue light ratio today of $\langle M/L_B\rangle_{g+s}>4.1\,I_{60}$ 
(the total blue luminosity density at the present--epoch is ${\cal L}_B=2\times 
10^8\,h\lden$, Ellis \etal 1996). The dependence of these estimates on the 
cosmological model (through eq. 4) is rather weak.

A visible mass density at the level of the above lower limit, while able to 
explain the measured sky brightness, requires all the stars that 
give origin to the observed EBL to have formed at very low redshifts
($z_F\lta 0.5)$, a scenario which appears to be at variance with the observed 
evolution of the UV luminosity density (Lilly \etal 1996; Madau \etal 1998).
For illustrative purposes, it is interesting to consider instead a model where
the star formation rate per unit comoving volume stays {\it constant} 
with cosmic time. In an EdS cosmology with $h=0.5$, one derives from equations
(2), (3), and (4) 
\begin{equation}
I_{\rm EBL}=1460\,\eblunits \langle {\dot\rho_s\over \sfrd}\rangle.
\end{equation}
The observed EBL therefore implies  a `fiducial' mean star 
formation rate (SFR) density of $\langle \dot\rho_s\rangle=0.04$ $I_{60}$ $\sfrd$,
a factor 3 {\it higher} than the value measured at 
$\langle z\rangle=0.15$ by Treyer \etal (1998).\footnote{For the Treyer \etal 
data we have used the 
conversion from UV luminosity density to SFR per unit volume,
$\log \dot \rho_s=\log {\cal L}_{2000}-28.1$, appropriate for the assumed IMF, 
and corrected upwards the observed ${\cal L}_{2000}$ by a factor of 1.8
for dust extinction.} Any value much smaller than this over a sizeable 
fraction of the Hubble time will generate an EBL well short of $60\,\eblunits$.
Ignoring for the moment the recycling of returned gas into new stars,
the {\it visible} mass density at the present epoch is simply   
$\rho_{g+s}=\int_0^{t_H} \dot \rho_s(t)dt= 5.2\times 10^8\,I_{60}\,\mden$, 
corresponding to $\Omega_{g+s}h^2=0.0019\,I_{60}$ and $\langle M/L_B\rangle_{g+s}
=5.2\,I_{60}$ (both values would be a factor of 1.7 higher in the case of a 
Salpeter IMF down to 0.1 $M_\odot$). 

A more realistic scenario which fits the most recent measurements of the rest--frame
UV--continuum and H$\alpha$  luminosity densities 
from the present--epoch to $z=4$ (after a correction for dust extinction 
is applied to the data) and produces a 
total EBL of about the right magnitude ($I_{60}=1$), is one where the SFR 
density {\it evolves} as (EdS, $h=0.5$)  
\begin{equation}
\dot\rho_s(z)={0.1\,e^{2.2z}\over e^{2.2z}+6}\,\sfrd. 
\end{equation}
This SFR density (an updated version of the one used in eq. 8 of MP) increases 
slowly -- by about a factor of 4 -- from the present-epoch 
to $z=1$ (cf. Cowie \etal 1999) and remains constant at $z>2$ (Steidel \etal 1999). 
Since about half of the present--day stars are formed at $z>1.5$ in this model and
their contribution to the EBL is redshifted away, the resulting visible mass density
is $\Omega_{g+s}h^2=0.0023\,I_{60}$ and $\langle M/L_B\rangle_{g+s}=6.3\,
I_{60}$, slightly larger than in the $\dot\rho_s=$const approximation. 

We conclude that, depending on the star formation history and for the assumed 
IMF, the observed EBL requires between 8\% and 12\% of the nucleosynthetic 
baryon density to be today in the forms of stars, recycled gas, and their 
remnants.
According to the most recent census of cosmic baryons, the mass density in 
stars and their remnants observed today is $\Omega_sh=0.00245^{+0.00125}_{-0.00088}$ (Fukugita \etal 1998), corresponding to a mean stellar mass--to--blue light 
ratio of $\langle M/L_B\rangle_s=3.4^{+1.7}_{-1.3}$ (roughly 70\% 
of this mass is found in old spheroidal populations). While this is lower than
the $\langle M/L_B\rangle_{g+s}$ ratio predicted by equation (6), one should note 
that efficient recycling of ejected gas into new star 
formation would tend to reduce the apparent difference in the budgets. 
With the adopted IMF, about 30\% of this mass will be returned to 
the interstellar medium in $10^8$ yr, after intermediate--mass stars eject 
their envelopes and massive stars explode as supernovae. This return fraction,
$R$, becomes 50\% after about 10 Gyr.\footnote{An asymptotic mass fraction 
of stars returned as gas, $R=\int (m-m_f)\phi(m)dm \times 
[\int m\phi(m)dm]^{-1}\approx 0.5$, can be obtained by using the semiempirical 
initial ($m$)--final ($m_f$) mass relation of Weidemann 
(1987) for stars with $1<m<10$, and by assuming
that stars with $m>10$ return all but a $1.4\msun$ remnant.}~ Alternatively, 
the gas returned by stars may be ejected into the intergalactic
medium. With an IMF-averaged yield of returned metals of $y_Z\approx 
1.5\,Z_\odot$,\footnote{Here we have taken $y_Z\equiv \int mp_{\rm zm}\phi(m)dm 
\times [\int m\phi(m)dm]^{-1}$, the stellar yields $p_{\rm zm}$ of Tsujimoto
\etal (1995), and a GBF$+$Salpeter 
IMF. In the case of a Salpeter IMF down to $0.1\,\msun$, $y_Z$ should be multiplied
by 0.6.}\, the predicted mean metallicity at the present epoch is 
$y_Z\Omega_{g+s}/\Omega_b\approx 0.2\,Z_\odot$, similar to the values inferred 
from cluster abundances (Renzini 1997).

\smallskip
\begin{table}
\small
\begin{center}
{TABLE 1}\\
\medskip
{\sc The Cosmic Stellar Baryon and Metal Budget}\\
\smallskip
\begin{tabular}{cccc}
\hline\hline
SF mode & $\Omega_{g+s}h^2$ & $\langle M/L_B\rangle_{g+s}$ & $y_Z\Omega_{g+s}/
\Omega_b$\\
\hline
instantaneous & 0.0015& 4.1&0.12 \\
constant & 0.0019& 5.2& 0.15\\
evolving & 0.0023&6.3 & 0.18\\
\hline
\end{tabular}
\end{center}
NOTE.---%
These values assume an EdS cosmology with $h=0.5$, a GBF$+$Salpeter IMF, and 
an EBL intensity of $60\,\eblunits$. 
\end{table}

The cosmic stellar baryon and metal budget is summarized in Table 1 for the three
different modes (instantaneous, constant, and evolving) of SF considered.
Note that a steeper IMF -- e.g. a Scalo function which is significantly less rich
in massive stars than Salpeter -- or an IMF which does not flatten 
below $0.6\,\msun$ would generate mass--to--light ratios that are too high 
compared to the observed values.

\section{EBL from MACHOs}

The nature of the dark matter in the halo of galaxies remains one of the
outstanding problems in astrophysics.
One of the most interesting constraints posed by the observed brightness of the
night sky concerns the possibility that a significant fraction of the dark mass in 
present--day galaxy halos may be associated with faint white--dwarf (WD) remnants 
of a population of intermediate--mass stars that formed at high redshifts. The 
latest results of the microlensing MACHO experiment 
towards the LMC, interpreted in the context of a Galactic dark matter halo 
consisting partially of compact objects,\footnote{Another widely discussed
possibility is that microlensing is actually dominated by self-lensing in a LMC 
halo.}\, give a MACHO halo fraction of 20\%, with a 95\% confidence interval of 
8--50\% (Alcock \etal 2000). Similar results have been reached by the EROS 
collaboration 
(Lasserre \etal 2000). The most likely MACHO mass is between 0.15 and 0.9
$\msun$: this mass scale is a 
natural one for WDs, a scenario also supported by the lack of a 
numerous spheroidal population of low--mass main sequence stars in the HDF 
(Gould \etal 1998), and by the recent discovery of two ancient halo WDs
in a systematic proper motion survey (Ibata \etal 2000).
The total mass of MACHOs inferred within 50 kpc is 
$9^{+4}_{-3}\times 10^{10}\,M_\odot$ (Alcock \etal 2000), implying a 
`MACHO--to--blue light' ratio for the Milky Way in the range 4 to 10 solar 
(cf Fields \etal 1998).
If these values were typical of the luminous universe as a whole, i.e. if 
MACHOs could be viewed as a new stellar population having similar 
properties in all disk galaxies, then the cosmological mass density 
of MACHOs today would be $\Omega_{\rm MACHO}=(4-10)\,f_B{\cal L}_B/
\rho_{\rm crit}=(0.003-0.007)\,f_B\,h^{-1}$, a significant entry in 
the cosmic baryon budget, $\Omega_{\rm MACHO}/\Omega_b=0.15-0.4\,f_Bh$.
Here $f_B\approx 0.5$ is 
the fraction of the blue luminosity density radiated by stellar disks (Fukugita
\etal 1998). 
Note that if MACHOs are halo WDs, the contribution of their {\it progenitors} 
to the mass density parameter is several times higher.

Halo IMFs which are very different from that of the solar
neighborhood, i.e. which are heavily--biased towards WD progenitors and 
have very few stars forming with masses below $2\,M_\odot$ (as these would
produce bright WDs in the halo today that are not seen) and above 
$8\,M_\odot$ (to avoid the overproduction of heavy elements), have been 
suggested as a suitable mechanism for explaining the microlensing data
(Chabrier \etal 1996). While the halo WD scenario 
may be tightly constrained by the observed rate of Type Ia SNe in galaxies 
(Smecker \& Wyse 1991), by the expected C and N overenrichment of halo stars 
(Gibson \& Mould 1997), and by the number counts of faint galaxies in deep 
optical surveys (Charlot \& Silk 1995), here we explore a potentially more 
direct method (as it is does not depend on, e.g. extrapolating stellar yields 
to primordial metallicities, on galactic winds removing the excess heavy 
elements 
into the intergalactic medium, or on the reddening of distant halos by dust),
namely we will compute the contribution of WD progenitors in dark galaxy halos 
to the extragalactic background light. 

Following Chabrier (1999) (see also MP), we adopt a truncated power--law IMF, 
\begin{equation}
\phi(m)={\rm const}\times e^{-(\overline{m}/m)^3}\, m^{-5}.
\end{equation}    
This form mimics a mass function strongly peaked 
at $0.84\,\overline{m}$. To examine the dependence of the IMF on 
the results we consider two functions (shown in Figure 2), $\overline{m}=2.4$ and 
$\overline{m}=4$: both yield a present--day Galactic halo mass--to--light ratio 
$>100$ after a Hubble time, as required in the absence of a large non--baryonic 
component. We further assume that a population of halo WD progenitors 
having mass density $X\Omega_bh^2=0.0193X$ formed 
instantaneously at redshift $z_F$ with this IMF and nearly primordial 
($Z=0.02\,Z_\odot$) metallicity. The resulting EBL from such an event is shown 
in Figure 4 for $X=0.1, 0.3,$ and 0.6 and a $\Lambda$--dominated 
universe with $\Omega_M=0.3$, $\Omega_\Lambda=0.7$, and $h=0.65$ ($t_H=14.5$ 
Gyr).

Consider the $\overline{m}=2.4$ case first. With $z_F=3$ and $X=0.6$, this 
scenario would generate an EBL at a level of $300\,\eblunits$.
Even if only 30\% of the nucleosynthetic baryons formed at $z_F=5$ with 
a WD--progenitor dominated IMF, the resulting background light at 
Earth would exceed the value of $100\,\eblunits$, the `best--guess' upper limit
to the observed EBL from the data plotted in Figure 1.
Since the return fraction of this IMF is $R\approx 0.8$, 
only 20\% of this stellar mass would be leftover as WDs, the rest 
being returned to the ISM. Therefore, if galaxy halos comprise 100\% 
of the nucleosynthetic baryons, only a small fraction of their mass, 
$X_{\rm WD}\approx 0.2\times 0.3=0.06$ could be in the form of WDs.\footnote{Pushing
 the peak of the IMF to more massive stars, $\overline{m}=
4$, helps only marginally. With $\overline{m}=2.4$, the energy radiated 
per stellar baryon over a timescale of 13 Gyr is equal to 2 MeV, 
corresponding to 10 MeV per baryon in WD remnants. A similar value is
obtained in the $\overline{m}=4$ case: because of the shorter lifetimes of 
more massive stars the expected EBL is reduced, but only by 20\% or so 
(see Figure 4). Moreover, the decreasing fraction of 
leftover WDs would raise more severe problems of metal galactic enrichment.}

In qualitative agreement with the most recent microlensing results (Alcock \etal
2000; Lasserre \etal 2000), these limits likely imply a non--baryonic dark halo.
On the other hand, we draw attention to the fact that, even if only 20\% of 
the baryons in the universe turned into halo WD progenitors at $z_F\gta 10$, 
their background 
light should be detectable as a peak in the EBL around $1\,\mu$m. In Figure 1 we 
show the EBL produced by such a WD--progenitor dominated IMF with $\overline{m}=4$ 
and $(z_F, X, X_{\rm WD})=(11, 0.2, 0.04)$, assuming negligible dust reddening. 
Intriguingly, this model appears to be consistent with measurements of 
the optical/near--IR EBL by Bernstein \etal (2000) and Gorjan \etal (2000). 

\section{EBL from quasar activity}

An accurate determination of the contribution of QSOs to the EBL, $I_{\rm QSO}$,
obtainable 
in principle by integrating the number--magnitude relation down to the 
detection threshold, is in practice made uncertain by our poor knowledge of the 
mean bolometric 
correction and of the sky brightness due to undetected faint objects. Such a direct
method also explicitly neglects the possible existence of a population of dusty 
AGNs with strong intrinsic absorption as invoked, e.g., in many 
models for the X--ray background (e.g. Madau \etal 1994; Comastri \etal 1995): these
Type II QSOs, while undetected at optical wavelengths, could contribute 
significantly to the far--IR background and to the total EBL. Moreover, the 
conversion of $I_{\rm QSO}$ 
to the expected mass density of quasar remnants at the present--epoch, 
$\rho_{\rm BH}$, (assuming a standard black hole accretion model for QSOs) 
requires a correction
for the radiation energy lost to the cosmic expansion which depends on the 
quasar redshift distribution.

Some (but hardly all) of these difficulties can perhaps be bypassed by computing 
instead, in analogy with the stellar case,
\begin{equation}
I_{\rm QSO}={c\over 4\pi}\int_0^{t_H}\, {dt\over 1+z} \int_0^\infty L\phi(L,t) 
dL, \label{eq:qso}
\end{equation}
where $\phi(L,t)$ is the comoving bolometric luminosity function of QSOs at cosmic 
time $t$.
We have modeled $\phi(L,t)$ adopting up--to-date determinations of the quasar number
density and spectral energy distribution from the far--IR to gamma--rays, and 
using new synthesis models for the cosmic X--ray background (XRB) based on a 
mixture of unabsorbed Type I and heavily absorbed Type II AGNs (Haardt \& Madau 
2000). The spectrum of the radiation background from QSOs 
is shown in Figure 5 (we indicate it with $J$ instead of $I$ as it has been 
filtered through the absorption of the intergalactic medium at UV frequencies, 
cf. Haardt \& Madau 1996) as a function of redshift. 
Our model can simultaneously reproduce the observed spectrum of the 
XRB and the most recent source counts in both the soft and hard X--ray bands, and
yields at the present--epoch a value of $I_{\rm QSO}\approx 2\,\eblunits$ (previous 
estimates range from 0.7 to 3 $\eblunits$, Soltan 1982; Chokshi \& Turner 1992; 
Small \& Blandford 1992). 
In determining the mass density of quasar remnants at the present--epoch we take 
into account
the large contribution due to the absorbed radiation of Type II AGNs (Fabian \&
Iwasawa 1999).  Such
radiation (which is not shown in the spectrum of Figure 5) is most
probably re--emitted in the far IR band. We derive
\begin{equation}
\rho_{\rm BH}={1\over c^2\eta}\int_0^{t_H} dt \int_0^\infty L\phi(L,t) dL\approx
3\times 10^4\,\eta^{-1}\,\mden,
\label{eq:mqso}
\end{equation}
where $\eta$ is the efficiency of accreted mass--to--radiation conversion, equal 
to 5.7\% for standard disk accretion onto a Schwarzschild black hole. 

It is in principle possible to gauge our model for the evolution and 
emission properties of the quasar population by  weighing the local mass 
density of black holes remnants. Recent dynamical evidence 
indicates that supermassive black holes reside at the center of most nearby 
galaxies (Richstone \etal 1998). The available data (more than 30 objects) show an
empirical correlation between bulge luminosity and black hole mass 
(Magorrian \etal 1998), which becomes even tighter when bulge velocity 
dispersion is plotted instead of luminosity (Ferrarese \& Merritt 2000; 
Gebhardt \etal 2000). The implied mean ratio of black hole to bulge mass is
$M_{\rm BH}\approx 0.0013 \,M_{\rm sph}$ as a best--fit (Merritt \& Ferrarese 2000),
a factor of 5 smaller than the mean ratio computed by Magorrian \etal (1998). 
The mass density in old spheroidal populations today is estimated to be
$\Omega_{\rm sph}h= 0.0018^{+0.0012}_{-0.00085}$ (Fukugita \etal 1998), 
implying a mean mass density of quasar remnants today of 
\begin{equation}
\rho_{\rm BH}\approx 6\times 10^5\,h\,\mden.
\end{equation}
This determination agrees with the result of equation (9)
for $\eta\sim 0.05\,h^{-1}$.
\bigskip\bigskip

\noindent We thank the organizer of the IAU Symposium 204, M. Harwit and M. Hauser,
for their patience in awaiting for this manuscript, and T. Matsumoto for 
discussions on MACHO progenitors and the EBL. Support for this work was provided 
by NASA through ATP grant NAG5--4236 (P. 
M.), and by C.N.A.A. and ASI through contract ASI-ARS-98-119 (L.P.). 

\references
Alcock, C., \etal 2000, ApJ, 542, 281

Bernstein, R. A., Freedman, W. L., \& Madore, B. F. 2000, submitted 

Bruzual, A. C., \& Charlot, S. 1993, ApJ, 405, 538  

Burles, S., \& Tytler, D. 1998, ApJ, 499, 699

Chabrier, G. 1999, ApJ, 513, L103

Chabrier, G., Segretain, L., \& Mera, D. 1996, ApJ, 468, L21

Charlot, S., \& Silk, J. 1995, ApJ, 445, 124

Chokshi, A., \& Turner, E. L. 1992, MNRAS, 259, 421

Comastri, A., \etal 1995, A\&A, 296, 1

Cowie, L. L., Songaila, A., \& Barger, A. J. 1999, AJ, 118, 603

Dwek, E., \& Arendt, R. G. 1998, ApJ, 508, L9
 
Dwek, E., \etal 1998, ApJ, 508, 106

Elbaz, D., \etal 1999, A\&A, 351, L37

Ellis, R.~S., Colless, M., Broadhurst, T., Heyl, J., \& Glazebrook,
K. 1996, MNRAS, 280, 235

Elmegreen, B. G. 1998, in Unsolved Problems in Stellar Evolution, ed. M. Livio
(Cambridge: Cambridge University Press), in press (astro--ph/9811289)

Fabian, A. C., \& Iwasawa, K. 1999, MNRAS, 303, L34

Ferrarese, L., \& Merritt, D. 2000, ApJ, 539, L9

Fields, B.D., Freese, K., \& Graff, D. S. 1998, NewA, 3, 347  

Finkbeiner, D. P., Davis, M., \& Schlegel, D. J. 2000, ApJ, submitted 
(astro--ph/0004175)

Fixsen, D. J., \etal 1998, ApJ, 508, 123

Fukugita, M., Hogan, C. J., \& Peebles, P. J. E. 1998, ApJ, 503, 518

Gardner, J.~P., Brown, T. M., \& Ferguson, H. C. 2000, ApJ, in press
(astro--ph/0008247)

Gebhardt, K., \etal 2000, ApJ, 539, L13
Gibson, B., \& Mould, J. 1997, ApJ, 482, 98

Gorjan, V., Wright, E. L., \& Chary, R. R. 2000, ApJ, 536, 500

Gould, A., Bahcall, J. N., \& Flynn, C. 1996, ApJ, 465, 759 (GBF)

Gould, A., Flynn, C., \& Bahcall, J. N. 1998, ApJ, 503, 798 

Haardt, F., \& Madau, P. 1996, ApJ, 461, 20 

Haardt, F., \& Madau, P. 2000, in preparation

Hauser, M. G., \etal 1998, ApJ, 508, 25

Ibata, R., Irwin, M., Bienaym\'e, O., Scholz, R., \& Guibert, J. 2000, ApJ,
532, L41
 
Lagache, G., Abergel, A., Boulanger, F., Desert, F. X., \& Puget, J.--L. 
1999, A\&A, 344, 322 

Lasserre, T.,\etal 2000, A\&A, 355, L39

Lilly, S.~J., Le F{\'e}vre, O., Hammer, F., \& Crampton,
D., 1996, ApJ, 460, L1

Madau, P., Ghisellini, G., \& Fabian, A. C. 1994, MNRAS, 270, L17

Madau, P., \& Pozzetti, L. 2000, MNRAS, 312, L9 (MP)

Madau, P., Pozzetti, L., \& Dickinson, M. 1998, ApJ, 498, 106 

Magorrian, G., \etal 1998, AJ, 115, 2285

Massey, P. 1998, in The Stellar Initial Mass Function, ed. G. Gilmore \& D. 
Howell (San Francisco: ASP), p. 17

Merritt, D., \& Ferrarese, L. 2000, preprint (astro-ph/0009076)

Paresce, F., \& De Marchi, G. 2000, ApJ, 534, 870

Renzini, A. 1997, ApJ, 488, 35 

Richstone, D., \etal 1998, Nature, 395, 14

Small, T. A., \& Blandford, R. D. 1992, MNRAS, 259, 725

Smecker, T. A., \& Wyse, R. 1991, ApJ, 372, 448 

Soltan, A. 1982, MNRAS, 200, 115

Steidel, C. C., Adelberger, K. L., Giavalisco, M., Dickinson, M., \&
Pettini, M. 1999, ApJ, 519, 1

Treyer, M. A., Ellis, R. S., Millard, B., Donas, J., \& Bridges, T. J.
1998, MNRAS, 300, 303

Tsujimoto, T., Nomoto, K., Yoshii, Y., Hashimoto, M., Yanagida, S.,
\& Thielemann, F.--K. 1995, MNRAS, 277, 945

Weidemann, V. 1987, A\&A, 188, 74

Zoccali, M., \etal 2000, ApJ, 530, 418

\begin{figure}
\centerline{
\psfig{figure=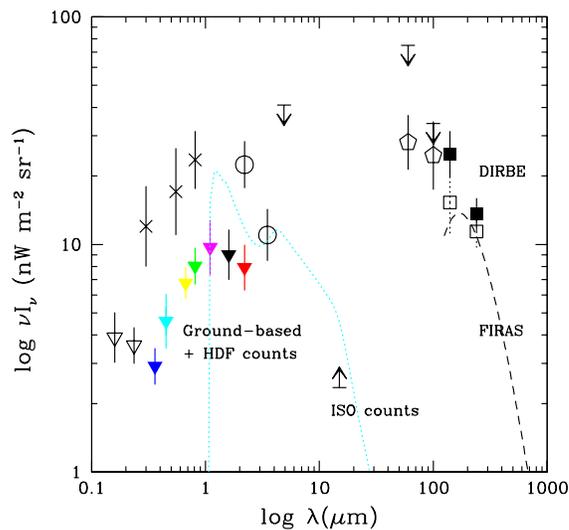,width=3.4in}}
\caption{\footnotesize Spectrum of the optical extragalactic 
background light from resolved sources 
as derived from a compilation of ground--based and space--based galaxy counts 
in the $UBVIJHK$ bands ({\it filled triangles}, MP), together with the FIRAS 
125--5000 $\mu$m ({\it dashed line}, Fixsen \etal 1998) and DIRBE 140 and 
240 $\mu$m ({\it filled squares}, Hauser \etal 1998) detections.
The {\it empty squares} show the DIRBE points after correction 
for WIM dust emission (Lagache \etal 1999). Also plotted ({\it empty 
triangles}) are the STIS NUV (2360 \AA) and FUV (1600 \AA) points from Gardner 
\etal (2000), together with the tentative detections at 2.2 and 3.5 $\mu$m 
({\it empty points}, Gorjan \etal 2000) and at 60 and 100 $\mu$m 
({\it empty pentagons}, Finkbeiner \etal 2000) 
from {\it COBE}/DIRBE observations.
The crosses at 3000, 5500, and 8000 \AA\ are Bernstein \etal (2000) 
measurements of the EBL from resolved and unresolved galaxies fainter 
than $V=23$ mag (the error bars showing 2$\sigma$ statistical errors).
Upper limits are from Hauser \etal (1998), the lower limit from Elbaz \etal 
(1999) ISO counts. The {\it dotted curve} shows the synthetic EBL produced by 
a WD--progenitor dominated IMF with 
$\overline{m}=4$ and $(z_F, X, X_{\rm WD})=(11, 0.2, 0.04)$, in the case of 
zero dust reddening.
}
\end{figure}

\begin{figure} 
\centerline{
\psfig{figure=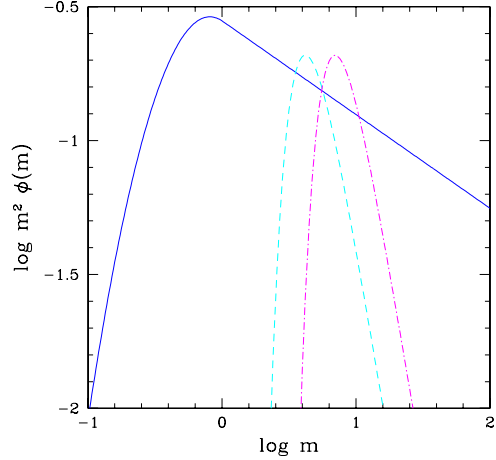,width=3.in}}
\caption{\footnotesize Stellar initial mass functions $\phi(m)$ (times $m^2$).
{\it Solid line:} Salpeter IMF, $\phi(m)\propto m^{-2.35}$ 
at high masses, matched to a GBF function at $m\le 1$. {\it Dotted line:} 
WD--progenitor dominated IMF in galaxy halos, $\phi(m)\propto 
e^{-(\overline{m}/ 
m)^3}\, m^{-5}$, with $\overline{m}=2.4$ (see text for details). 
{\it Dot--dashed line:} Same for $\overline{m}=4$.
All IMFs have been normalized to $\int m\phi(m)dm=1$.
}
\end{figure}

\begin{figure}
\plottwo{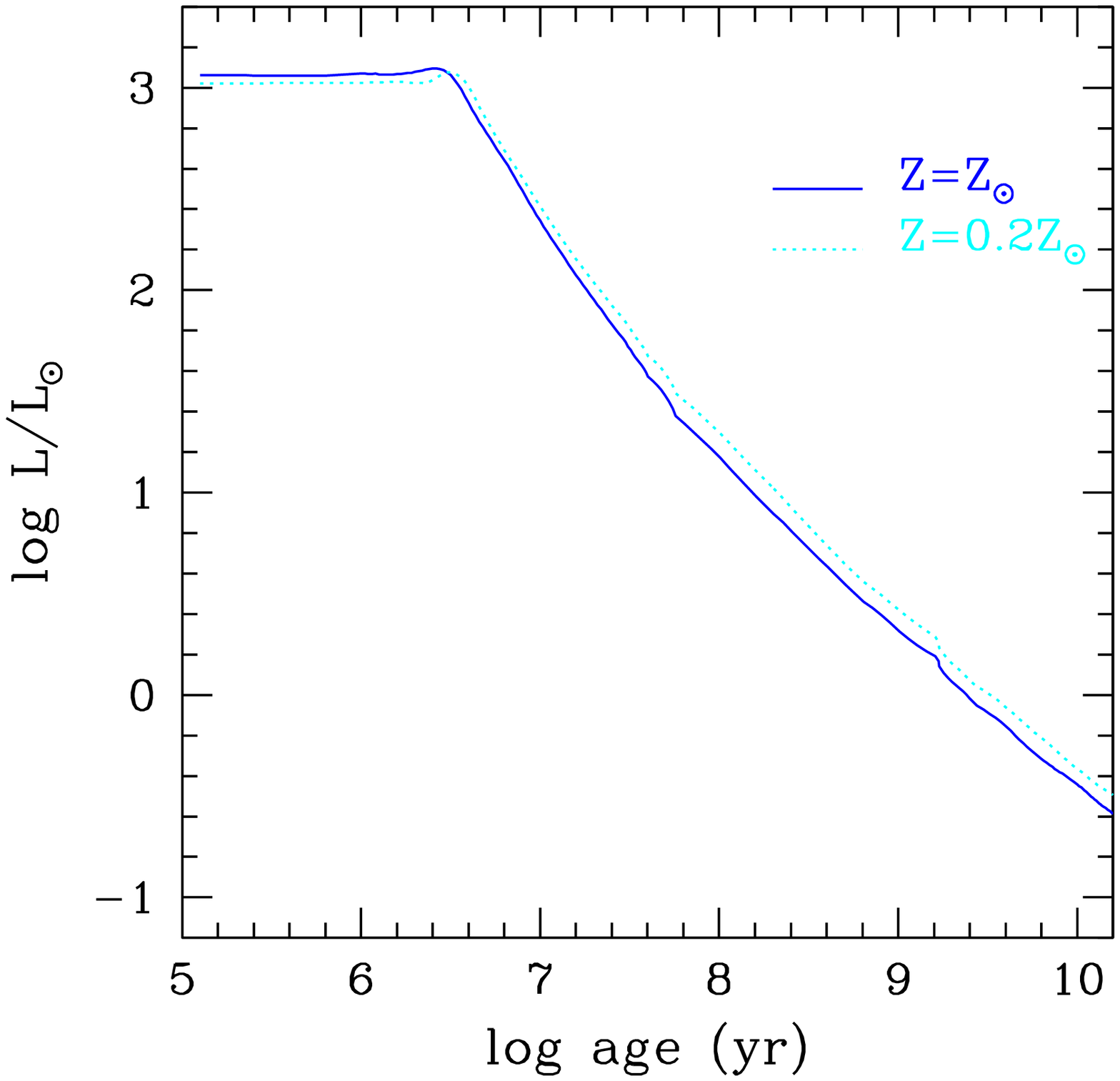}{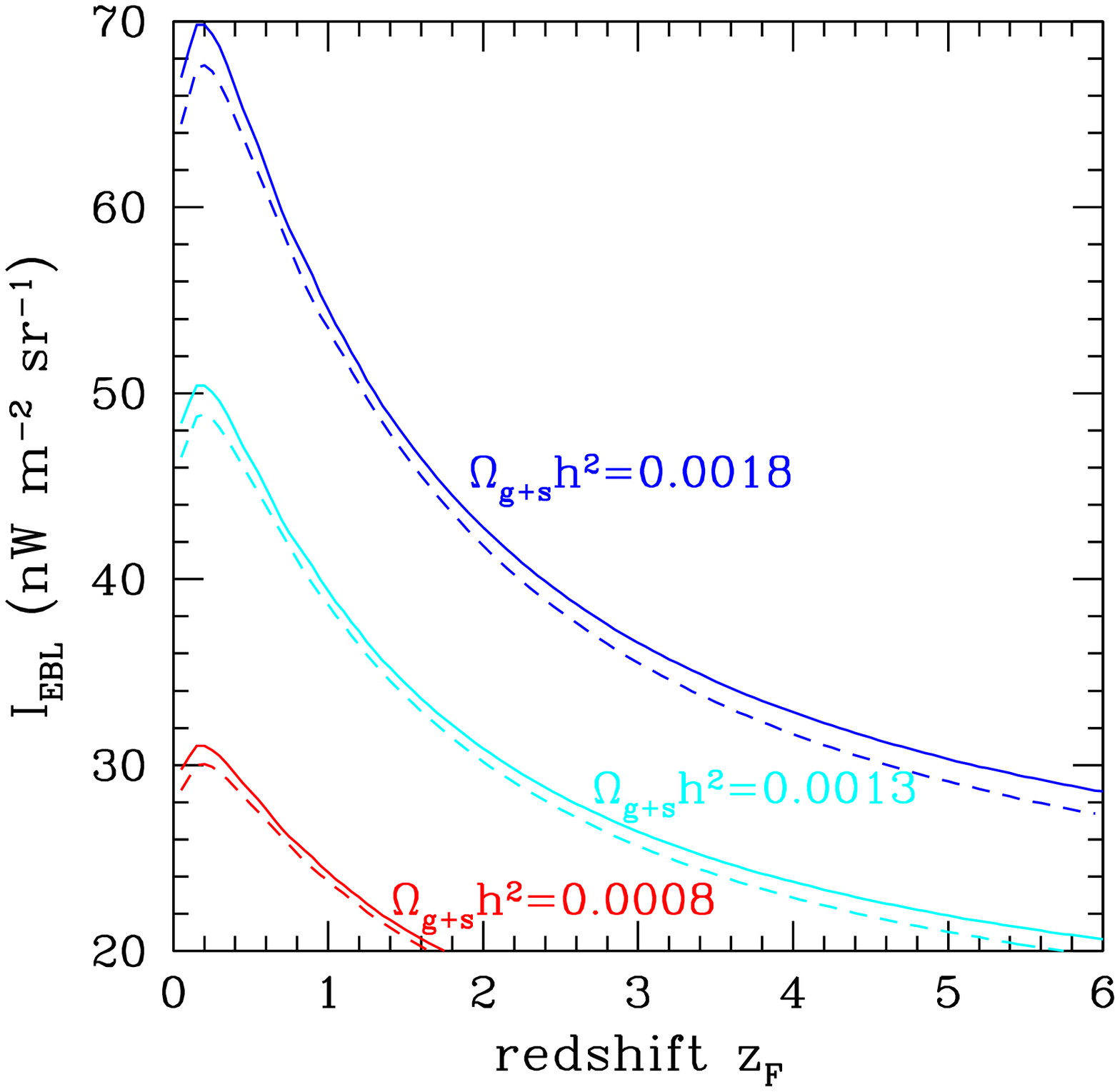}
\caption{\footnotesize {\it Left:} Synthetic (based on an update of 
Bruzual \& Charlot's 1993 libraries)
bolometric luminosity versus age of a simple stellar 
population having total mass $M=1\,M_\odot$, metallicity $Z=Z_\odot$
({\it solid line}) and $Z=0.2\,Z_\odot$ ({\it dotted line}), and 
a GBF$+$Salpeter IMF (see text for details). 
{\it Right:} EBL observed at Earth from the instantaneous formation at 
redshift $z_F$ of the same stellar population ($Z=Z_\odot$ case) with
total mass density $\Omega_{g+s}h^2=0.0018, 0.0013,$ and 0.0008.
{\it Solid curves:} EdS universe with $h=0.5$ ($t_H=13$ Gyr). 
{\it Dashed curves:}  
$\Lambda$--dominated universe with $\Omega_M=0.3$, $\Omega_\Lambda=0.7$, and
$h=0.65$ ($t_H=14.5$ Gyr).
}
\end{figure}

\begin{figure}
\plottwo{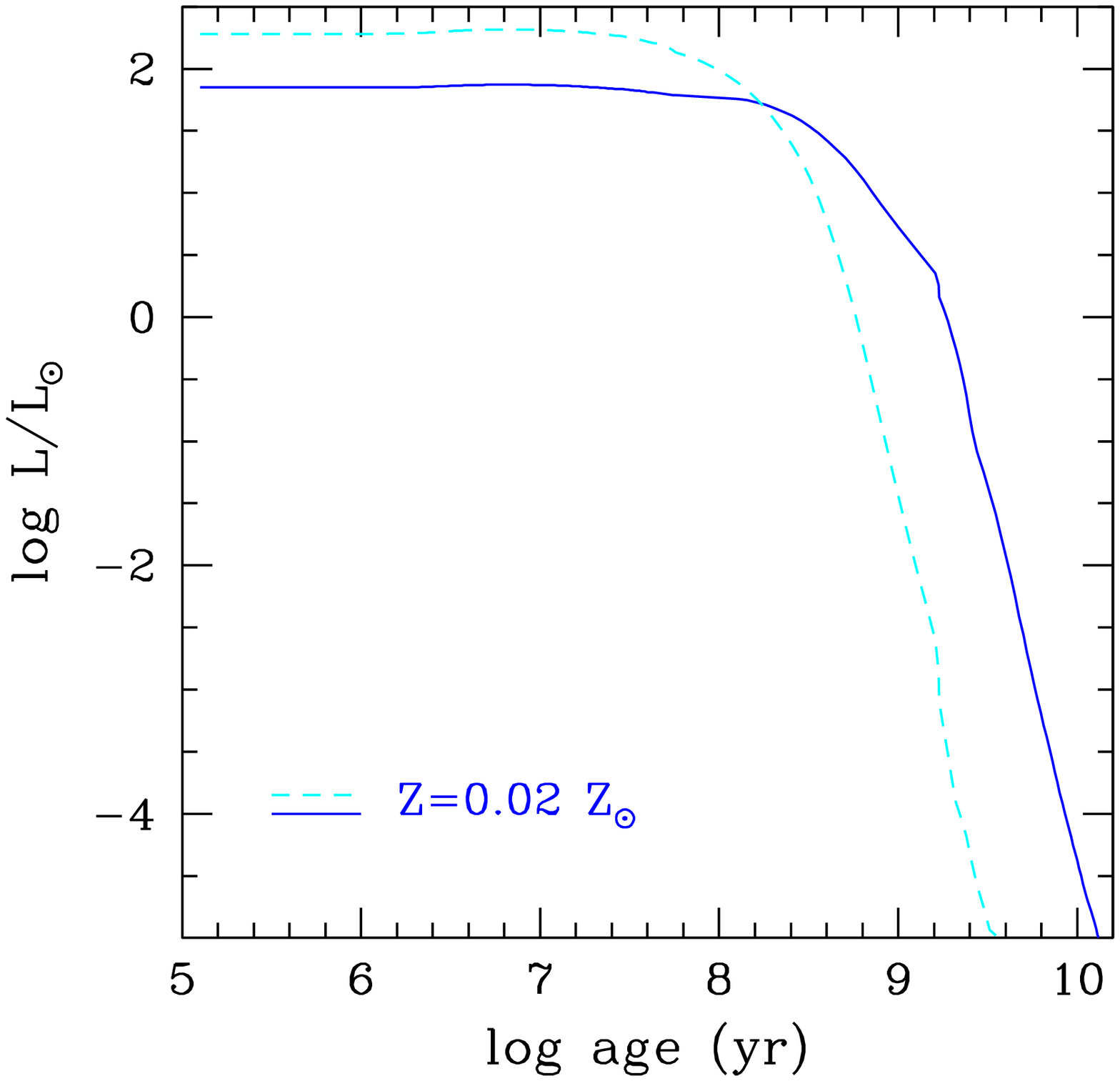}{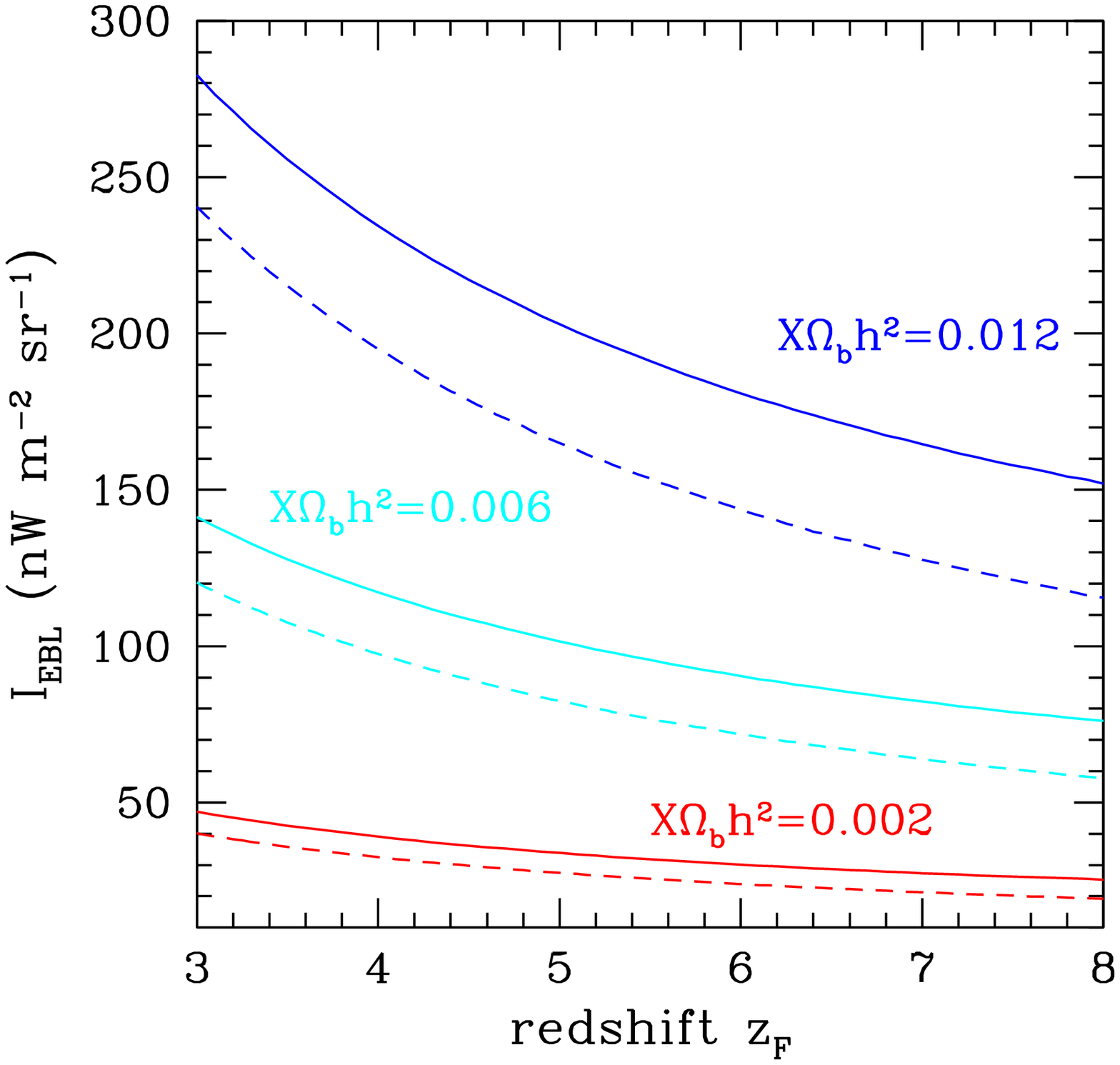}
\caption{\footnotesize {\it Left:} Synthetic bolometric luminosity versus age of a simple 
stellar population having total mass $M=1\,M_\odot$, metallicity $Z=0.02
Z_\odot$, and a WD--progenitor dominated IMF (see text for details) 
with $\overline{m}=2.4$
({\it solid line}) and $\overline{m}=4$ ({\it dashed line}). 
{\it Right:} EBL observed at Earth from the instantaneous formation at 
redshift $z_F$ of a stellar population having the same IMF and metallicity, and 
mass density $X\Omega_bh^2=0.012, 0.006$ and 0.002 (corresponding to 60, 30, 
and 10 per cent of the nucleosynthetic value of Burles \& Tytler 1998), as 
a function of  
$z_F$. A $\Lambda$--dominated universe with $\Omega_M=0.3$, $\Omega_\Lambda=
0.7$, and $h=0.65$ has been assumed. {\it Solid line}:
$\overline{m}=2.4$. {\it Dashed line}: $\overline{m}=4$. 
}
\end{figure}

\begin{figure} 
\plottwo{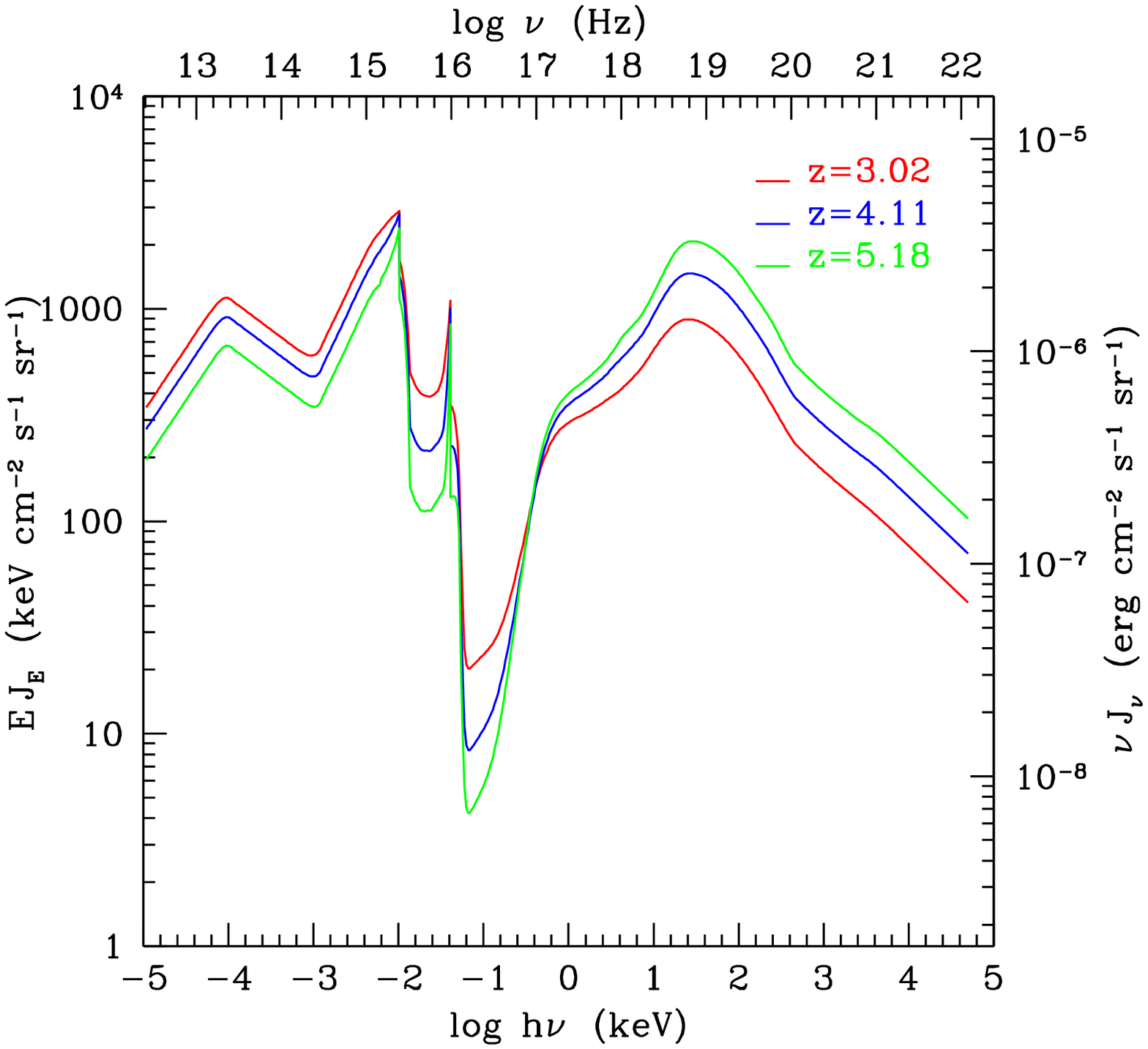}{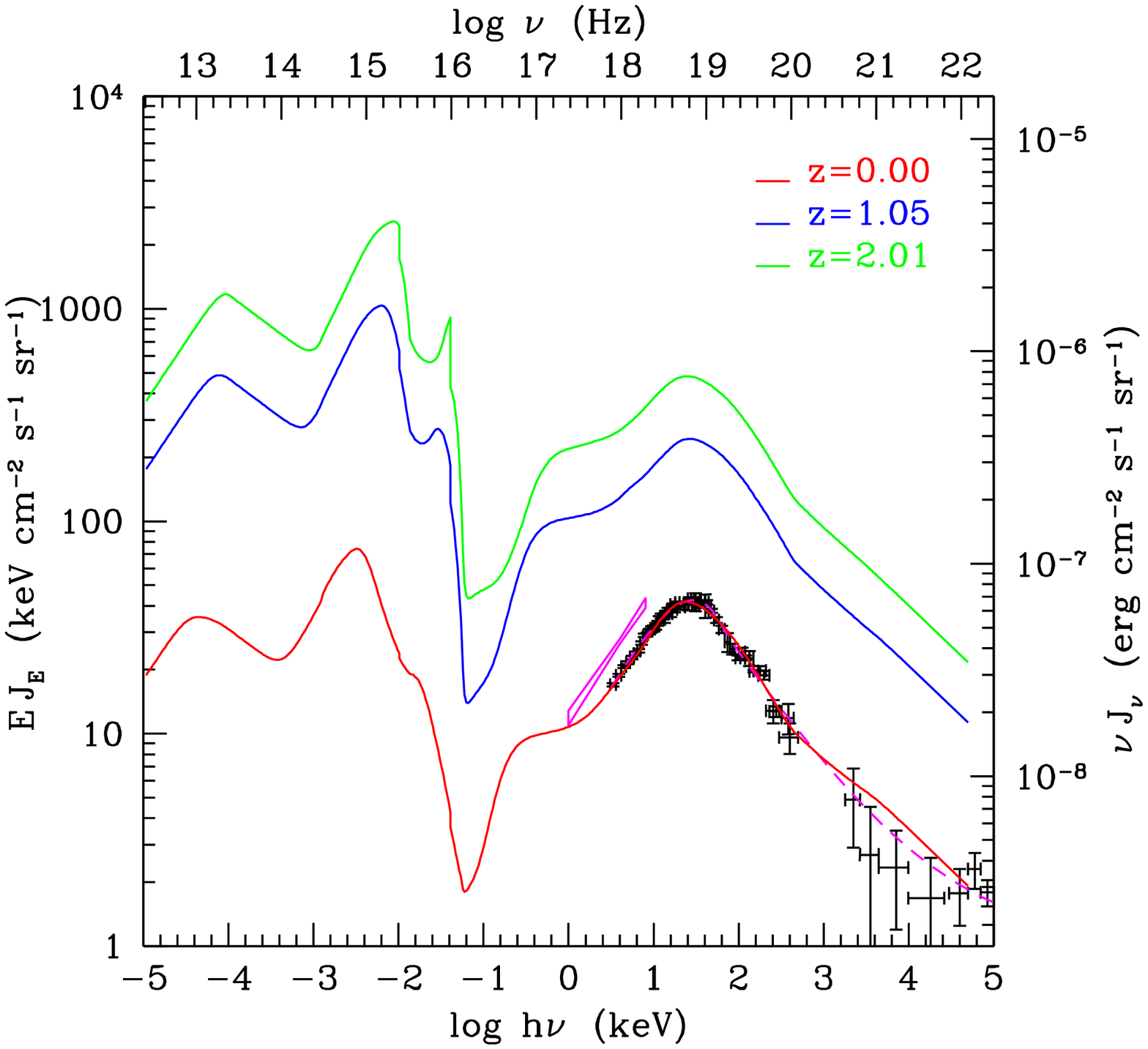}
\caption{\footnotesize Synthetic radiation background from quasars including 
the reprocessing of UV radiation from intervening absorption systems. The data 
points at $z=0$ show the spectrum of the observed X and $\gamma-$ray background.
(From Haardt \& Madau 2000.)} 
\end{figure}

\end{document}